\documentclass[showpacs,preprintnumbers,amsmath,amssymb]{revtex4}

\usepackage{graphicx}
\usepackage{bm}
\usepackage{epstopdf}
\usepackage{float}
\usepackage{dcolumn}
\usepackage{bm}
\usepackage{mathrsfs}
\usepackage{amsmath}

\begin{document}

\title{Semi-commuting and commuting operators for the Heun family}
\author{D. Batic}
\email{dbatic@pi.ac.ae}
\affiliation{%
Department of Mathematics,\\  Khalifa University of Science and Technology,\\ PI Campus, Abu Dhabi, United Arab Emirates}
\author{D. Mills}
\email{dominic.millz27@gmail.com}
\affiliation{
Department of Mathematics,\\
University of the West Indies,\\
Mona Campus, Kingston, Jamaica
}
\author{M. Nowakowski}
\email{mnowakos@uniandes.edu.co}
\affiliation{
Departamento de Fisica,\\ Universidad de los Andes, Cra.1E
No.18A-10, Bogota, Colombia
}

\date{\today}

\begin{abstract}
\noindent
We derive the most general families of differential operators of first and second degree semi-commuting with the differential operators of the Heun class. Among these families we classify all those families commuting with the Heun class. In particular, we discover that a certain generalized Heun equation commutes with the Heun differential operator allowing us to construct the general solution to a complicated fourth order linear differential equation with variable coefficients which Maple $16$ cannot solve.   
\end{abstract}

\pacs{Valid PACS appear here}
\maketitle
\noindent
\textbf{Keywords}: semi-commuting operators, commuting operators,  Heun equation, confluent Heun equation, biconfluent Heun equation, double confluent Heun equation, triconfluent Heun equation, generalized Heun equation, factorizations
\newpage
\pagenumbering{arabic}
\section{Introduction}
\noindent
The interest in commuting differential expressions started almost $140$ years ago with the work of Floquet followed by contributions of Wallenberg and Schur \cite{FWS} but the decisive input came from the work of Burchnall and Chaundy which allowed to connect this mathematical area to algebraic geometry \cite{BC123}. Since many operators in mathematical physics do not commute, the most celebrated example being represented by the position and momentum operators in Quantum Mechanics, the theory of non-commuting operators is also interesting in its own. Furthermore, there can be cases when certain symmetries or degeneracies are present leading differential operators to {\it{almost}} commute, or {\it{semi-commute}} but what does it mean that two operators semi-commute? Let us consider two operators $P$ and $T$ of degree $m$ and $n$, respectively. The commutator $[P,T]$ will be in general an operator of degree $m+n-1$. Let us suppose that $P$ is given and $Q$ be an arbitrary operator of degree $n$. We say that $P$ and $Q$ semi-commute if the highest order term in $[P,Q]$ vanish, i.e. the commutator of these two operators is an operator of degree $m+n-2$. The problem of finding operators with analytic coefficients that semi-commute with a given monic differential operator having analytic coefficients has been thoroughly treated in \cite{Gorder}. The interest in the construction of a class of operators semi-commuting with a given operator $P$ associated to some equation of mathematical physics relies in the fact that such a class may contain a subclass of commuting operators. Hence, starting with a certain differential operator and after having computed the corresponding class of semi-commuting operators, we can impose additional constraints that if fulfilled they can lead to a class of commuting operators. If two operators commute, then the Burchnall and Chaundy theory ensures that they must share a solution which in turn we may use to reduce the order of the ODE associated to the operator $P$. In what follows we consider a given differential operator $P$ with analytic coefficients, namely
\begin{equation}\label{P}
P=d^2_x+p_1(x) d_x+p_0(x),\quad d_x:=\frac{d}{dx}.
\end{equation}
The operator
\[
Q=q_n(x)d^n_x+q_{n-1}(x)d^{n-1}_x+\cdots+q_1(x)d_x+q_0(x)
\]
semi-commute with $P$ if we can find coefficients $q_0,\cdots,q_n$ such that \cite{Gorder}
\begin{eqnarray}
d_x q_n&=&0,\label{C1}\\
d_x q_k&=&\frac{1}{2}\sum_{j=k+1}^n\binom{j}{k}q_j(x)d_x^{j-k}p_1(x)-\frac{1}{2}P[q_{k+1}(x)]\quad\forall k=0,1,\cdots,n-1.\label{C2}
\end{eqnarray}
If the additional condition 
\begin{equation}\label{C3}
d^2_x q_0(x)+p_1(x)d_xq_0(x)=\sum_{k=1}^n q_k(x)d_x^k p_0(x)
\end{equation}
is satisfied for all $x$, then $P$ and $Q$ commute. Note that in principle the unknown coefficients $q_k$'s can be obtained from (\ref{C2}) and will depend on some integration constants $\beta_i$. Substituting these solutions in (\ref{C3}) yield the relation
\begin{equation}\label{C4}
\varphi(x;\beta_0,\beta_1,\cdots,\beta_n)=0.
\end{equation}
In some cases the $\beta_i$'s can be chosen so as to permit commutativity. If this is not the case, we can look instead for a solution $x_0=x_0(\beta_0,\beta_1,\cdots,\beta_n)$ to (\ref{C4}). If such a solution exists, we say that $P$ and $Q$ {\it{locally commute}} at $x=x_0$. In this work we extend the treatment of Airy, Bessel, and hypergeometric operators presented in \cite{Gorder} to differential operators associated to the Heun equation and its confluent forms \cite{Ronveaux,Slavyanov} displayed in Table~\ref{Heun}. We recall that the Heun equation(HE) has been originally constructed by the German mathematician Karl Heun (1889) \cite{Heun} as a generalization of the hypergeometric equation. To underline the importance of the HE in mathematical physics we recall that it contains the generalized spheroidal equation, the Coulomb spheroidal equation, Lam´e, Mathieau, and Ince equations as special
cases. The fields of applications of the HE in physics are so large that it is not possible to describe them here in detail. However, a review of many general situations relevant to physics, chemistry, and engineering where the HE and its confluent forms occur can be found in \cite{Ronveaux,Slavyanov}. Since we will show in the next section that a certain generalized  Heun  equation (GHE)  commutes  with  the  Heun  differential  operator, thus  allowing  us  to  construct the general solution of a rather complicated fourth order linear ODE  with variable coefficients, some comments on the GHE are in order. The GHE is a second order differential equation with three regular singular points and one irregular singular point at infinity. It generalizes the ellipsoidal wave equation as well as the Heun equation. \cite{SS} obtained under certain conditions all connection coefficients between the Floquet solutions at the finite singularities, thus determining the full monodromy group of the GHE. This equation plays an important role in applications in the context of Quantum Field Theory in curved space-times. More precisely, 
it describes the radial spinors of an electron or neutrino immersed in the gravitational potential of a Kerr-Newman black hole and also the static perturbations for the non-extremal Reissner-Nordstr\"{o}m
solution to the Einstein field equations \cite{BSW}.\\
Before we derive the classes of semi-commuting and commuting operators of first and second degree, we show that  equation (\ref{C2}) derived in \cite{Gorder} is not correct. This can be already seen in the case $n=1$. Let $P$ be given as in (\ref{P}). Suppose that we want to construct the class of operators $Q$ of degree one commuting with $P$. Let $Q=\beta_1 d_x+q_0(x)$ with $q_0$ a yet-to-be determined function. It is not difficult to verify that
\[
[Q,P]=[2d_x q_0(x)-\beta_1 d_x p_1(x)]d_x+d^2_x q_0(x)+p_1(x)d_x q_0(x)-\beta_1d_x p_0(x).
\]
Hence, the function $q_0$ will be represented by the solution of the first order ODE
\begin{equation}\label{COR1}
d_x q_0(x)=\frac{\beta_1}{2}d_x p_1(x),
\end{equation}
whereas the commutativity condition reads
\begin{equation}\label{COR2}
d^2_x q_0(x)+p_1(x)d_x q_0(x)-\beta_1d_x p_0(x)=0.
\end{equation}
However, even though in the case $n=1$ equation (\ref{C3}) coincides with our (\ref{COR2}), equation (\ref{C2}) becomes instead
\[
d_x q_0(x)=\frac{\beta_1}{2}d_x p_1(x)-\frac{\beta_1}{2}p_0(x)
\]
and as a result $Q$ and $P$ cannot commute. Hence, $(27)$ in \cite{Gorder} should be taken with some caution. In the case $n=2$ we consider the same operator $P$ as before but now $Q=\beta_2 d_{xx}+q_1(x) d_x+q_0(x)$. It can be easily verified that the operators $Q$ and $P$ will commute whenever
\begin{eqnarray}
d_x q_1(x)&=&\beta_2 d_x p_1(x),\label{COR3}\\
d_x q_0(x)&=&\frac{1}{2}q_1(x) d_x p_1(x)+\frac{\beta_2}{2}d^2_x p_1(x)+\beta_2 d_xp_0(x)-\frac{1}{2}\left[d^2_x+p_1(x) d_x\right]q_1(x), \label{COR4}\\
d^2_x q_0(x)+p_1(x) d_x q_0(x)&=&q_1(x) d_x p_0(x)+\beta_2 d^2_x p_0(x) \label{COR5}.
\end{eqnarray}
Notice again that equation (\ref{C3}) for $n=2$ reproduces correctly our equation (\ref{COR5}) but (\ref{C2}) or equivalently  $(27)$ in \cite{Gorder} leads to the wrong results  
\[
d_x q_1(x)=\beta_2 d_xp_1(x)-\frac{\beta_2}{2} p_0(x),\quad
d_xq_0(x)=\frac{1}{2}q_1(x) d_x p_1(x)+\frac{1}{2}\beta_2 d^2_x p_1(x)-\frac{1}{2}P[q_1(x)].
\]
\begin{table}[h]
\centering
\caption{Heun family of differential operators $P$ as defined in (\ref{P})}
\label{Heun}
\begin{tabular}{llll}
\hline
Equation &  $p_1(x)$ & $p_0(x)$ & parameters \\ \hline
Heun     &  $\frac{\gamma}{x}+\frac{\delta}{x-1}+\frac{\epsilon}{x-a}$   & $\frac{\alpha\beta x-q}{x(x-1)(x-a)}$   &   $\epsilon=\alpha+\beta+1-\delta-\gamma,\quad\alpha,\beta,\gamma,\delta,q\in\mathbb{C},\quad a\in\mathbb{R}\backslash\{0,1\}$         \\
confluent Heun       & $p+\frac{\gamma}{x}+\frac{\delta}{x-1}$    &  $\frac{p\alpha x-q}{x(x-1)}$  &  $p,q,\alpha,\gamma,\delta\in\mathbb{C}$   \\       
reduced confluent Heun      & $\frac{\gamma}{x}+\frac{\delta}{x-1}$    &  $\frac{\kappa x+q}{x(x-1)}$  & $\kappa,\gamma,\delta,q\in\mathbb{C}$  \\        biconfluent Heun &$\frac{\tau}{x}+\frac{\nu}{x^2}-1$ &$-\frac{\alpha x+q}{x}$ & $\tau,\nu,\alpha,q\in\mathbb{C}$ \\
double confluent Heun &$\frac{\tau}{x}+\frac{\nu}{x^2}-1$ & $-\frac{\alpha x+q}{x^2}$&$\tau,\nu,\alpha,q\in\mathbb{C}$\\
triconfluent Heun &$\sigma-x^2$ &$\alpha x-q$ & $\sigma,\alpha,q\in\mathbb{C}$\\
representative triconfluent Heun & $0$ & $A_0+A_1 x+A_2 x^2-\frac{9}{4} x^4$& $A_0,A_1,A_2\in\mathbb{C}$
\\ \hline
\end{tabular}
\end{table}

\section{Semi-commuting operators for the Heun family}
\noindent
We construct classes of semi-commuting operators for the differential operators associated to different Heun-like differential equations. We also investigate the problem of the existence of subclasses of commuting operators among the aforementioned classes. For each differential operator associated to the equations presented in Table~\ref{P} we derive when possible all operators of first and second degree commuting with it. 

\subsection{The Heun operator}
\subsubsection{The case $n=1$}\label{n1HH}
With the help of (\ref{COR1}) we find the following family of first degree operators   
\begin{equation}\label{QH1}
Q=\beta_1 d_x+\frac{\beta_1}{2}\left(\frac{\gamma}{x}+\frac{\delta}{x-1}+\frac{\epsilon}{x-a}\right)+\beta_0,\quad 
\epsilon=\alpha+\beta+1-\delta-\gamma
\end{equation}
semi-commuting with the operator $P_{H}$ associated to the Heun equation. The commutativity condition (\ref{COR2}) will be satisfied for certain choices of the parameters that we list here below. Note that if $L:=QP_{H}=P_{H}Q$, the solutions to the equations $Q\varphi=0$ and $P_{H}\psi=0$ allow to construct the general solution to the third order ODE $Lf=0$ with
\begin{equation}\label{LHn1}
L=\beta_1 d^3_x+\left(\rho+\sum_{i=1}^3\frac{A_i}{x-x_i}\right)d^2_x+\left(\sum_{i=1}^3\frac{B_i}{x-x_i}\right)d_x+\sum_{i=1}^3\frac{C_i}{x-x_i}
\end{equation}
where $x_1=0$, $x_2=1$, and $x_3=a$.
\begin{enumerate}
\item
Case $\alpha=\gamma=\delta=q=0$, and $\beta=1$ or $\beta=\gamma=\delta=q=0$, and $\alpha=1$. The general solutions to the ODEs $P_H\psi=0$ and $Q\varphi=0$ are
$\psi(x)=c_1+c_2(x-a)^{-1}$ and $\varphi(x)=c_3(x-a)^{-1}\mbox{exp}(-\beta_0 x/\beta_1)$, respectively. Moreover, the general solution to the ODE $Lf=0$ with $L$ as in (\ref{LHn1}), $\rho=\beta_0$, $A_3=3\beta_1$ and $B_3=2\rho$, while all other coefficients are zero, is given by
\[
f(x)=c_1+\frac{1}{x-a}\left(c_2+c_3e^{-\frac{\beta_0}{\beta_1}x}\right).
\]
\item
Case $\alpha=\gamma=q=0$, $\beta=1$ and $\delta=2$ or $\beta=\gamma=q=0$, $\alpha=1$, and $\delta=2$. See case $1.$ with $a=1$. If $\alpha=\delta=q=0$, $\beta=1$, and $\gamma=2$ or $\beta=\delta=q=0$, $\alpha=1$, and $\gamma=2$, see case $1.$ with $a=0$.
\item
Case $\alpha=\gamma=\delta=q=0$, and $\beta=-1$ or $\beta=\gamma=\delta=q=0$, and $\alpha=-1$. The general solutions to the ODEs $P_H\psi=0$ and $Q\varphi=0$ are $\psi(x)=c_1+c_2 x$ and $\varphi(x)=c_3 \mbox{exp}(-\beta_0 x/\beta_1)$, respectively.
Moreover, the general solution to the ODE $Lf=0$ with $L$ as in (\ref{LHn1}), $\rho=\beta_0$, and $A_i=B_i=C_i=0$ for all $i=1,2,3$ is
\[
f(x)=c_1+c_2 x+c_3e^{-\frac{\beta_0}{\beta_1}x}.
\]
\item
Case $\gamma=q=0$, $\alpha=\delta=2$, and $\beta=1$ or $\gamma=q=0$, $\beta=\delta=2$, and $\alpha=1$. The general solutions to the ODEs $P_H\psi=0$ and $Q\varphi=0$ are $\psi(x)=(x-1)^{-1}(x-a)^{-1}(c_1+c_2 x)$ and  $\varphi(x)=(x-1)^{-1}(x-a)^{-1}\mbox{exp}(-\beta_0 x/\beta_1)$, respectively. Moreover, the general solution to the ODE $Lf=0$ with $L$ as in (\ref{LHn1}), $A_1=B_1=C_1=0$ 
\[
\rho=\beta_0,\quad A_2=A_3=3\beta_1,\quad
B_2=2\beta_0-\frac{6\beta_1}{a-1},\quad B_3=4\beta_0-B_2,\quad
C_2=-\frac{2\beta_0}{a-1},\quad C_3=-C_2
\]
is 
\[
f(x)=\frac{1}{(x-1)(x-a)}\left(c_1+c_2 x+c_3e^{-\frac{\beta_0}{\beta_1}x}\right).
\]
\item
Case $\beta=\gamma=\delta=2$, $\alpha=3$, and $q=2a+2$ or $\alpha=\gamma=\delta=2$, $\beta=3$, and $q=2a+2$. The general solutions to the ODEs $P_H\psi=0$ and $Q\varphi=0$ are $\psi(x)=[x(x-1)(x-a)]^{-1}(c_1+c_2 x)$ and $\varphi(x)=c_3[x(x-1)(x-a)]^{-1}\mbox{exp}(-\beta_0 x/\beta_1)
$, respectively. Furthermore, the general solution to the ODE $Lf=0$ with $L$ as in (\ref{LHn1}), $\rho=\beta_0$, $A_i=3\beta_1$ for all $i=1,2,3$ and
\begin{eqnarray*}
B_1&=&2\beta_0-\frac{6\beta_1}{a}(a-1),\quad
B_2=2\beta_0+\frac{6\beta_1(a-2)}{a-1},\quad
B_3=2\beta_0+\frac{6\beta_1(2a-1)}{a(a-1)},\\
C_1&=&\frac{6\beta_1-2\beta_0(a+1)}{a},\quad
C_2=\frac{2\beta_0(a-2)-6\beta_1}{a-1},\quad
C_3=\frac{2\beta_0(2a-1)+6\beta_1}{a(a-1)}
\end{eqnarray*}
is given by
\[
f(x)=\frac{1}{x(x-1)(x-a)}\left(c_1+c_2 x+c_3e^{-\frac{\beta_0}{\beta_1}x}\right).
\]
\item
Case $\alpha=\gamma=q=2$, $\beta=1$, and $\delta=0$ or $\beta=\gamma=q=2$, $\alpha=1$, and $\delta=0$. The general solutions to the ODEs $P_H\psi=0$ and $Q\varphi=0$ are  $\psi(x)=[x(x-a)]^{-1}(c_1+c_2 x)$ and  $\varphi(x)=c_3[x(x-a)]^{-1}\mbox{exp}(-\beta_0 x/\beta_1)$. Moreover, the general solution to the ODE $Lf=0$ with $L$ as in (\ref{LHn1}), $\rho=\beta_0$, $A_2=B_2=C_2=0$ and
\[
A_1=A_3=3\beta_1,\quad
B_1=2\beta_0-\frac{6\beta_1}{a},\quad 
B_3=4\beta_0-B_1,\quad 
C_1=-\frac{2\beta_0}{a},\quad C_3=-C_1
\]
is 
\[
f(x)=\frac{1}{x(x-a)}\left(c_1+c_2 x+c_3e^{-\frac{\beta_0}{\beta_1}x}\right).
\]
\end{enumerate}

\subsubsection{The case $n=2$}\label{HeunN2}
With the help of (\ref{COR3}) and (\ref{COR4}) we find the following family of second degree operators   
\begin{equation}\label{QHn2}
Q=\beta_2 d^2_x+\left[\beta_2\left(\frac{\gamma}{x}+\frac{\delta}{x-1}+\frac{\epsilon}{x-a}\right)+\beta_1\right] d_x+\frac{a_3 x^3+a_2 x^2+a_1 x+a_0}{x(x-1)(x-a)}
\end{equation}
with $\epsilon=\alpha+\beta+1-\delta-\gamma$ and
\[
a_3=\beta_0,\quad a_2=-\beta_0(a+1)+\frac{\beta_1}{2}(\delta+\epsilon+\gamma),\quad
a_1=\beta_0 a+\alpha\beta\beta_2-\frac{\beta_1}{2}\left[a(\delta+\gamma)+\epsilon+\gamma\right],\quad
a_0=\frac{1}{2}\beta_1\gamma a-\beta_2 q
\]
semi-commuting with the operator $P_{H}$. The commutativity condition (\ref{COR5}) will be satisfied for certain choices of the parameters that we list here below. Observe that if $L:=QP_{H}=P_{H}Q$, the general solution to the fourth order ODE $Lf=0$ with
\begin{equation}\label{LHn2}
L=d^4_x+\left(\nu+\sum_{i=1}^3\frac{\mathfrak{A}_i}{x-x_i}\right)d^3_x+\left[\mu+\sum_{k=1}^2\sum_{i=1}^3\frac{\mathfrak{B}_{i,k}}{(x-x_i)^k}\right]d^2_x+\left(\sum_{k,i=1}^3\frac{\mathfrak{C}_{i,k}}{(x-x_i)^k}\right)d_x+\sum_{k,i=1}^3\frac{\mathfrak{D}_{i,k}}{(x-x_i)^k}
\end{equation}
can be immediately constructed from the solutions of the equations $Q\varphi=0$ and $P_{H}\psi=0$.
\begin{enumerate}
\item
Case $\beta_1=0$, and $\beta_2\neq 0$. The operator $P_H$ is represented by the general Heun operator given in Table~\ref{Heun} and $Q$ is given by (\ref{QHn2}) with $a_3=\mu$, $a_2=-\mu(a+1)$, $a_1=\mu a+\alpha\beta$, $a_0=-q$ and $\mu=\beta_0/\beta_2$. Moreover, the general solution to the ODE $P_H\psi=0$ is
\[
\psi(x)=c_1 H(a,q,\alpha,\beta,\gamma,\delta;x)+c_2 x^{1-\gamma}H(a,q-(\gamma-1)(a\delta+\epsilon),\beta-\gamma+1,\alpha-\gamma+1,2-\gamma,\delta;x),
\]
where $H(\cdot)$ denotes the Heun function. Regarding the equation $Q\varphi=0$ observe that it has an irregular singular point at infinity and three finite regular singular points. The transformation $\varphi(x)=e^{Ax}f(x)$ with $A^2=-a_3$ brings the aforementioned equation into the generalized Heun equation (GHE)\cite{BSW}
\[
d^2_x f+\left(\frac{\gamma}{x}+\frac{\delta}{x-1}+\frac{\epsilon}{x-a}+\kappa\right)d_x f+\frac{b_2 x^2+b_1 x+b_0}{x(x-1)(x-a)}f=0
\]
with $\kappa=2A$, $b_0=Aa\gamma+a_0$, $b_1=Aa(A-\delta-\gamma)-A(\epsilon+\gamma)+a_1$, $b_2=-A^2(a+1)+A(\alpha+\beta+1)+a_2$. Then, the general solution to the equation $Q\varphi=0$ in a neighbourhood of the singularities $x_i$ of the GHE can be written as $\phi_{i}(x)=e^{Ax}\left[c_{3,i} h_{1,i}(x)+c_{4,i} h_{2,i}(x)\right]$, where $h_{1,i}$ and $h_{2,i}$ are two particular solutions to the GHE defined up to the next singularity. Then, the general solution to the ODE $Lf=0$ with $L$ as in (\ref{LHn2}), $\nu=0$ and
\begin{eqnarray*}
\mathfrak{A}_1&=&2\gamma,\quad \mathfrak{A}_2=2\delta,\quad \mathfrak{A}_3=2\epsilon,\quad
\mathfrak{B}_{1,1}=-\frac{2}{a}\left[(a\delta+\epsilon)\gamma+q\right],\quad
\mathfrak{B}_{2,1}=\frac{2}{a-1}\left[\gamma\delta(a-1)-\alpha\beta-\delta\epsilon+q\right],\\
\mathfrak{B}_{3,1}&=&\frac{2}{a(a-1)}\left[a(\alpha\beta+\delta\epsilon+\gamma\epsilon(a-1)-q)\right],\quad \mathfrak{B}_{1,2}=\gamma(\gamma-2),\quad \mathfrak{B}_{2,2}=\delta(\delta-2),\quad \mathfrak{B}_{3,2}=\epsilon(\epsilon-2),\\
\mathfrak{C}_{1,1}&=&\frac{1}{a^2}\left[\mu\gamma a^2+2a(\alpha\beta\gamma+\delta q-\gamma q)+2q(\epsilon-\gamma)\right],\\
\mathfrak{C}_{2,1}&=&\frac{1}{(a-1)^2}\left[\mu\delta(a-1)^2+2a(\gamma q-\delta q-\alpha\beta\gamma)+2\alpha\beta(\gamma+\epsilon-\delta)+2q(2\delta-\epsilon-\gamma)\right],\\
\mathfrak{C}_{3,1}&=&\frac{1}{a^2(a-1)^2}\left[\mu\epsilon a^2(a-1)^2+2\alpha\beta a^2(\gamma+\delta-\epsilon)-2a(\alpha\beta\gamma+\delta q-2\epsilon q+\gamma q)+2q(\gamma-\epsilon)\right],
\end{eqnarray*}
\begin{eqnarray*}
\mathfrak{C}_{1,2}&=&\frac{1}{a}\left[\gamma(\delta a+\epsilon)+2q(1-\gamma)\right],\quad
\mathfrak{C}_{2,2}=-\frac{1}{a-1}\left[\gamma\delta(a-1)+2(\alpha\beta-q)(\delta-1)-\delta\epsilon\right],\\
\mathfrak{C}_{3,2}&=&\frac{1}{a(a-1)}\left[2(\alpha\beta a-q)(\epsilon-1)-\epsilon a(\gamma+\delta)+\gamma\epsilon\right],\quad \mathfrak{C}_{i,3}=-\mathfrak{B}_{i,2}~\forall i=1,2,3,\\
\mathfrak{D}_{1,1}&=&\frac{1}{a^3}\left[\gamma a(a+1)(\alpha\beta a-q)-a^2 q(\delta+\mu)-2aq(\alpha\beta+q)+q(2q-\epsilon-\gamma)\right],\\
\mathfrak{D}_{2,1}&=&\frac{1}{(a-1)^3}\left[(a-1)^2(\alpha\beta\gamma-\alpha\beta\mu+\gamma q+\delta q+\mu q)-q(a-2)(\delta+2q)+\alpha\beta(2aq-6q+2\alpha\beta-\delta-\epsilon)\right],\\
\mathfrak{D}_{3,1}&=&\frac{1}{a^3(a-1)^3}\left[(a-1)^2((\mu a^2+\gamma)(\alpha\beta a-q)-\gamma q)+a^2(\alpha\beta a-3q)(\delta+\epsilon-2\alpha\beta)+q(a(3\epsilon-2\alpha\beta)+2q-\epsilon)\right],\\
\mathfrak{D}_{1,2}&=&\frac{q}{a^2}(q-\delta a-\epsilon),\quad
\mathfrak{D}_{2,2}=\frac{1}{(a-1)^2}\left[(\alpha\beta-q)^2+(\alpha\beta-q)(\gamma a-\gamma-\epsilon)\right],\\
\mathfrak{D}_{3,2}&=&\frac{1}{a^2(a-1)^2}\left[\alpha\beta a^2(\alpha\beta-\delta-q)+\alpha\beta a(\gamma-2q)+aq(\delta+\gamma)+q(q-\gamma)\right],\quad 
\mathfrak{D}_{1,3}=\frac{q}{a}(\gamma-2),\\
\mathfrak{D}_{2,3}&=&\frac{1}{a-1}(\alpha\beta-q)(\delta-2),\quad \mathfrak{D}_{3,3}=-\frac{(\alpha\beta a-q)(\epsilon-2)}{a(a-1)}
\end{eqnarray*}
is 
\begin{eqnarray*}
f_i(x)&=&c_1 H(a,q,\alpha,\beta,\gamma,\delta;x)+c_2 x^{1-\gamma}H(a,q-(\gamma-1)(a\delta+\epsilon),\beta-\gamma+1,\alpha-\gamma+1,2-\gamma,\delta;x)+\\
&&e^{Ax}\left[c_{3,i} h_{1,i}(x)+c_{4,i} f_{h,i}(x)\right]
\end{eqnarray*}
with $i=1,2,3$. It is interesting to observe that Maple $16$ fails to solve the fourth order ODE $Lf=0$. Furthermore, note that $Q=\beta_2 P_H$ whenever $\beta_0=0$. In this case the general solution to the ODE $Lf=0$ is expressed in terms of Heun functions only. 
\item 
Case $\alpha=\gamma=\delta=q=0$ and $\beta=1$, or $\beta=\gamma=\delta=q=0$, and $\alpha=1$. The operator $P_H$ and the solution to $P_H\psi=0$ are given as in case $1.$ ($n=1$) of the Heun operator. Moreover, the solution to $Q\varphi=0$ is 
\begin{equation}\label{mpmCH}
\varphi(x)=\frac{1}{x-a}\left(c_3 e^{m_{-}x}+c_4 e^{m_+ x}\right),\quad
m_\pm=-\frac{1}{2\beta_2}\left(\beta_1\pm\sqrt{\beta_1^2-4\beta_0\beta_2}\right).
\end{equation}
Then, the general solution to the ODE $Lf=0$ with $L$ as in (\ref{LHn2}) and non-vanishing coefficients $\mu=\beta_0/\beta_2$, $\nu=\beta_1/\beta_2$, $\mathfrak{A}_3=4$, $\mathfrak{B}_{3,1}=3\nu$ and $\mathfrak{C}_{3,1}=2\mu$
is
\[
f(x)=c_1+\frac{1}{x-a}\left(c_2+c_3 e^{m_{-}x}+c_4 e^{m_+ x}\right).
\]
\item
Case $\alpha=\gamma=q=0$, $\beta=1$, and $\delta=2$, or $\beta=\gamma=q=0$, $\alpha=1$, and $\delta=2$.  We find that $P_H$ and the solution to $P_H\psi=0$ are given as in case $2.$ ($n=1$) of the Heun operator. Moreover, $Q$, the solution to $Q\varphi=0$, the operator $L$, and the solution of $Lf=0$ can be obtained from the previous case with $a=1$. If instead $\alpha=\delta=q=0$, $\beta=1$, and $\gamma=2$, or $\beta=\delta=q=0$, $\alpha=1$, and $\gamma=2$,  we find that $P_H$ and the solution to $P_H\psi=0$ are given as in case $1.$ ($n=1$) of the Heun operator with $a=0$. Moreover, $Q$, the solution to $Q\varphi=0$, the operator $L$, and the solution to $Lf=0$ can be obtained from case $2.$ ($n=2$) of the Heun operator with $a=0$.
\item
Case $\alpha=\gamma=\delta=q=0$, and $\beta=-1$, or $\beta=\gamma=\delta=q=0$, and $\alpha=-1$. The operator $P_H$ and the solution to $P_H\psi=0$ are given as in case $3.$ ($n=1$) of the Heun operator. Furthermore, $Q=\beta_2 d^2_x+\beta_1 d_x+\beta_0$ and the solution to $Q\varphi=0$ is  $\varphi(x)=c_3 e^{m_- x}+c_4 e^{m_+ x}$ with $m_\pm$ given in (\ref{mpmCH}). Then, the general solution to the equation $Lf=0$ with $L=\beta_2 d^4_x+\beta_1 d^3_x+\beta_0 d_x$ is
\[
f(x)=c_1+c_2 x+c_3 e^{m_- x}+c_4 e^{m_+ x}.
\]
\item
Case $\gamma=q=0$, $\alpha=\delta=2$, and $\beta=1$, or $\gamma=q=0$, $\beta=\delta=2$, and $\alpha=1$. We find that $P_H$ and the solution of $P_H\psi=0$ are given as in case $4.$ ($n=1$) of the Heun operator. Moreover, the solution to $Q\varphi=0$ is given by $\varphi(x)=[(x-1)(x-a)]^{-1}\left[c_3 e^{m_- x}+c_4 e^{m_+ x}\right]$ with $m_\pm$ as in (\ref{mpmCH}). Then, the general solution to the ODE $Lf=0$ with $L$ as in (\ref{LHn2}) and non-vanishing coefficients $\mu=\beta_0/\beta_2$, $\nu=\beta_1/\beta_2$ and
\[
\mathfrak{A}_2=\mathfrak{A}_3=4,\quad\mathfrak{B}_{2,1}=3\nu-\frac{12}{a-1},\quad \mathfrak{B}_{3,1}=6\nu-\mathfrak{B}_{2,1},\quad
\mathfrak{C}_{2,1}=2\mu-\frac{6\nu}{a-1},\quad\mathfrak{C}_{3,1}=4\mu-\mathfrak{C}_{2,1}
\]
is
\[
f(x)=\frac{1}{(x-1)(x-a)}\left[c_1+c_2 x+c_3 e^{m_- x}+c_4 e^{m_+ x}\right].
\]
\item
Case $\beta=\gamma=\delta=2$, $\alpha=3$, and $q=2a+2$, or $\alpha=\gamma=\delta=2$, $\beta=3$, and $q=2a+2$. The solutions to the equations $P_H\psi=0$ and $Q\varphi=0$ are $\psi(x)=[x(x-1)(x-a)]^{-1}(c_1+c_2 x)$ and $\varphi(x)=[x(x-1)(x-a)]^{-1}\left(c_3 e^{m_- x}+c_4 e^{m_+ x}\right)$, respectively. Then, the general solution to the ODE $Lf=0$ with $L$ as in (\ref{LHn2}) and non-vanishing coefficients $\mu=\beta_0/\beta_2$, $\nu=\beta_1/\beta_2$, $\mathfrak{A}_1=\mathfrak{A}_2=\mathfrak{A}_3=4$ and
\begin{eqnarray*}
\mathfrak{B}_{1,1}&=&3\nu-\frac{12(a+1)}{a},\quad
\mathfrak{B}_{2,1}=3\nu+\frac{12(a-2)}{a-1},\quad
\mathfrak{B}_{3,1}=3\nu+\frac{12(2a-1)}{a(a-1)},\\
\mathfrak{C}_{1,1}&=&2\mu-\frac{6\nu(a+1)-24}{a},\quad
\mathfrak{C}_{2,1}=2\mu+\frac{6\nu(a-2)-24}{a-1},\quad
\mathfrak{C}_{3,1}=2\mu+\frac{6\nu(2a-1)+24}{a(a-1)},\\
\mathfrak{D}_{1,1}&=&\frac{2(3\nu-\mu)-2\mu}{a},\quad
\mathfrak{D}_{2,1}=\frac{2\mu(2a-1)-6\nu}{a-1},\quad
\mathfrak{D}_{3,1}=\frac{2\mu(2a-1)+6\nu}{a(a-1)}
\end{eqnarray*}
is
\[
f(x)=\frac{1}{x(x-1)(x-a)}\left(c_1+c_2 x+c_3 e^{m_- x}+c_4 e^{m_+ x}\right).
\]
\item
Case $\alpha=\gamma=q=2$, $\beta=1$, and $\delta=0$, or $\beta=\gamma=q=2$, $\alpha=1$, and $\delta=0$. The operator $P_H$ and the solution to $P_H\psi=0$ are given as in $7.$ ($n=1$) of the Heun operator. Moreover, the solution to $Q\varphi=0$ is $\phi(x)=[x(x-a)]^{-1}\left(c_3 e^{m_- x}+c_4 e^{m_+ x}\right)$. 
Finally, the solution to the ODE $Lf=0$ with $L$ as in (\ref{LHn2}) and non-vanishing coefficients $\mu=\beta_0/\beta_2$, $\nu=\beta_1/\beta_2$, $\mathfrak{A}_1=\mathfrak{A}_3=4$ and
\[
\mathfrak{B}_{1,1}=3\nu-\frac{12}{a},\quad
\mathfrak{B}_{3,1}=6\nu-\mathfrak{B}_{1,1},\quad
\mathfrak{C}_{1,1}=2\mu-\frac{6\nu}{a},\quad
\mathfrak{C}_{3,1}=4\mu=\mathfrak{C}_{1,1},\quad
\mathfrak{D}_{1,1}=-\frac{2\mu}{a},\quad
\mathfrak{D}_{3,1}=-\mathfrak{D}_{1,1}
\]
is represented by
\[
f(x)=\frac{1}{x(x-a)}\left(c_1+c_2 x+c_3 e^{m_- x}+c_4 e^{m_+ x}\right).
\]
\end{enumerate}

\subsection{The confluent Heun operator}
\subsubsection{The case $n=1$}\label{cHn1}
Using (\ref{COR1}) yields the following family of first degree operators   
\begin{equation}\label{QCH1}
Q=\beta_1 d_x+\frac{\beta_1}{2}\left(\frac{\gamma}{x}+\frac{\delta}{x-1}\right)+\beta_0
\end{equation}
semi-commuting with the operator $P_{CH}$ associated to the confluent Heun equation. The commutativity condition (\ref{COR2}) will be satisfied for certain choices of the parameters that we list here below. Note that if $L:=QP_{CH}=P_{CH}Q$, the solutions to the equations $Q\varphi=0$ and $P_{CH}\psi=0$ allow to construct the general solution to the third order ODE $Lf=0$ with
\begin{equation}\label{LCHn1}
L=\beta_1 d^3_x+\left(\xi+\sum_{i=1}^2\frac{\mathfrak{a}_i}{x-x_i}\right)d^2_x+\left(\eta+\sum_{i=1}^2\frac{\mathfrak{b}_i}{x-x_i}\right)d_x+\sum_{i=1}^2\frac{\mathfrak{c}_i}{x-x_i}
\end{equation}
where $x_1=0$ and $x_2=1$.
\begin{enumerate}
\item
Cases $\gamma=p=q=\delta=0$; $\gamma=p=q=0$ and $\delta=2$;   $p=q=\delta=0$ and $\gamma=2$; $\gamma=\delta=2$, $p=0$ and $q=-2$  have been already analyzed in Section~\ref{n1HH}.
\item
Case $\alpha=\gamma=q=\delta=0$. The solution to $P_H\psi=0$ is $\psi(x)=c_1+c_2 e^{-px}$. Furthermore, $Q$ is given as in case $3.$ ($n=1$) of the Heun operator. Finally, the solution of the third order ODE $Lf=0$ with $L=\beta_1 d^3_x+(\beta_1 p+\beta_0)d^2_x+\beta_0 p d_x$ is
\[
f(x)=c_1+c_2 e^{-px}+c_3 e^{-\frac{\beta_0}{\beta_1}x} .
\]
\item
Case $\alpha=1$, $\gamma=q=0$ and $\delta=2$. The solution to the equation $P_{CH}\psi=0$ is $\psi(x)=(x-1)^{-1}\left(c_1+c_2e^{-px}\right)$. Furthermore, $Q$ and the solution to $Q\varphi=0$ can be obtained from Section~\ref{n1HH}. The general solution to the equation $Lf=0$ with $L$ as in (\ref{LCHn1}) and non-vanishing coefficients $\xi=\beta_1 p+\beta_0$, $\eta=\beta_0 p$, $\mathfrak{a}_2=3\beta_1$, $\mathfrak{b}_2=2\xi$, $\mathfrak{c}_2=\eta$ 
is
\[
f(x)=\frac{1}{x-1}\left(c_1+c_2e^{-px}+c_3 e^{-\frac{\beta_0}{\beta_1}x}\right).
\]
\item
Case $\alpha=1$, $\gamma=2$, $q=p$ and $\delta=0$. The solution to the equation $P_{CH}\psi=0$ is $\psi(x)=x^{-1}\left(c_1+c_2e^{-px}\right)$. Moreover, $Q$ and the solution to $Q\varphi=0$ can be obtained from Section~\ref{n1HH}. The general solution to the equation $Lf=0$ with $L$ as in (\ref{LCHn1}) and non-vanishing coefficients $\xi=\beta_1 p+\beta_0$, $\eta=\beta_0 p$, $\mathfrak{a}_1=3\beta_1$, $\mathfrak{b}_1=2\xi$, $\mathfrak{c}_1=\eta$ is 
\[
f(x)=\frac{1}{x}\left(c_1+c_2e^{-px}+c_3 e^{-\frac{\beta_0}{\beta_1}x}\right).
\]
\item
Case $\alpha=\gamma=\delta=2$ and $q=p-2$. The solution to $P_{CH}\psi=0$ is $\psi(x)=[x(x-1)]^{-1}\left(c_1+c_2 e^{-px}\right)$. Moreover, $Q$ and the solution to $Q\varphi=0$ can be obtained from Section~\ref{n1HH}. The general solution to the equation $Lf=0$ with $L$ as in (\ref{LCHn1}) and non-vanishing coefficients $\xi=\beta_1 p+\beta_0$, $\eta=\beta_0 p$, $\mathfrak{a}_1=\mathfrak{a}_2=3\beta_1$, $\mathfrak{b}_1=\xi-\mathfrak{a}_1$, $\mathfrak{b}_2=2\xi-\mathfrak{b}_1$, $\mathfrak{c}_1=\eta-2\xi$, $\mathfrak{c}_2=2\eta-\mathfrak{c}_1$ 
is
\[
f(x)=\frac{1}{x(x-1)}\left(c_1+c_2 e^{-px}+c_3 e^{-\frac{\beta_0}{\beta_1}x}\right).
\]
\end{enumerate}
In the case of the reduced confluent Heun differential operator $P_{RCH}$ we find that the most general first degree differential operator $Q$ semi-commuting with $P_{RCH}$ is given by (\ref{QCH1}). By means of (\ref{COR2}) we can verify that the operators $P_{RCH}$ and $Q$ will commute whenever $\gamma=k=q=\delta=0$, or $\gamma=k=q=0$, and $\delta=2$, or $\gamma=2$, and $k=q=\delta=0$, or $\gamma=q=\delta=2$, and $k=0$  but these cases reduce to one of the cases treated above.

\subsection{The case $n=2$}
By means of (\ref{COR3}) and (\ref{COR4}) we find the following family of second degree operators   
\[
Q=\beta_2 d^2_x+\left[\beta_2\left(\frac{\gamma}{x}+\frac{\delta}{x-1}\right)+\beta_1\right]d_x+\frac{\beta_1\gamma+\beta_2(2q-\gamma p)}{x}+\frac{\beta_1\delta+\beta_2(2\alpha p-\delta p-2q)}{x-1}+\beta_0
\]
semi-commuting with the operator $P_{CH}$ associated to the confluent Heun equation. The commutativity condition (\ref{COR5}) will be satisfied for certain choices of the parameters that we list here below. Notice that if $L:=QP_{CH}=P_{CH}Q$, the solutions to the equations $Q\varphi=0$ and $P_{CH}\psi=0$ allow to construct the general solution the the fourth order ODE $Lf=0$ with
\begin{equation}\label{LCHn2}
L=\beta_2d^4_x+\left(\Gamma+\sum_{i=1}^2\frac{\mathfrak{A}_i}{x-x_i}\right)d^3_x+\left[\lambda+\sum_{k,i=1}^2\frac{\mathfrak{B}_{i,k}}{(x-x_i)^k}\right]d^2_x+\left[\eta+\sum_{k=1}^3\sum_{i=1}^2\frac{\mathfrak{C}_{i,k}}{(x-x_i)^k}\right]d_x+\sum_{k=1}^3\sum_{i=1}^2\frac{\mathfrak{D}_{i,k}}{(x-x_i)^k}
\end{equation}
with $\eta=\beta_0 p$.
\begin{enumerate}
\item
Case $\beta_1=\beta_2 p$. The solution to $P_{CH}\psi=0$ is \cite{Ronveaux}
\[
\psi(x)=c_1 H_c(a_1,a_2,a_3,a_4,a_5;x)+c_2 x^{1-\gamma}H_c(a_1,-a_2,a_3,a_4,a_5;x),
\]
with
\[
a_1=p,\quad a_2=\gamma-1,\quad a_3=\delta-1,\quad a_4=\frac{p}{2}(2\alpha-\gamma-\delta),\quad a_5=\frac{\gamma}{2}(p-\delta)-q+\frac{1}{2}.
\]
Moreover, the solution to $Q\varphi=0$ is given by
\[
\varphi(x)=e^{\widetilde{a}x}\left[c_3H_c(\widetilde{a}+a_1,a_2,a_3,a_4,a_5;x)+c_4 x^{1-\gamma}H_c(\widetilde{a}+a_1,-a_2,a_3,a_4,a_5;x)\right]
\]
with
\[
\widetilde{a}=\frac{1}{2}\left(-p+\sqrt{\frac{\beta_2 p^2-4\beta_0}{\beta_2}}\right).
\]
Then, the general solution to the equation $Lf=0$ with $L$ as in (\ref{LCHn2}) and non-vanishing coefficients  $\Gamma=p$, $\mathfrak{A}_1=\gamma$, $\mathfrak{A}_2=\delta$ and
\[
\lambda=\beta_2 p^2+\beta_0,\quad
\mathfrak{B}_{1,1}=2\beta_2(q+\gamma p-\gamma\delta),\quad
\mathfrak{B}_{2,1}=2\beta_2(\alpha p+\gamma\delta+p\delta-q),\quad
\mathfrak{B}_{1,2}=\beta_2\gamma(\gamma-2),\quad
\mathfrak{B}_{2,2}=\beta_2\delta(\delta-2),
\]
\[
\mathfrak{C}_{1,1}=\beta_0\gamma+2\beta_2(\gamma q+pq-\delta q-\alpha\gamma p),\quad
\mathfrak{C}_{2,1}=\beta_0\delta+2\beta_2(\alpha\gamma p+\alpha p^2+\delta q-\gamma q),\quad
\mathfrak{C}_{1,2}=\beta_2(\gamma\delta-\gamma p+2\gamma q-2q),
\]
\begin{eqnarray*}
\mathfrak{C}_{2,2}&=&\beta_2(2\alpha\delta p-2\alpha p-\gamma\delta-\delta p-2\delta q+2q),\quad 
\mathfrak{C}_{1,3}=-\mathfrak{B}_{1,2},\quad
\mathfrak{C}_{2,3}=-\mathfrak{B}_{2,2},\\
\mathfrak{D}_{1,1}&=&\beta_0 q+\beta_2(2q^2+\gamma q+\delta q-2\alpha pq-\alpha\gamma p),\quad
\mathfrak{D}_{2,1}=\beta_2[\alpha\gamma p+q(2\alpha p-\delta-\gamma-2q)]+\beta_0(\alpha p-q),\\
\mathfrak{D}_{1,2}&=&\beta_2\gamma(\delta-p+q),\quad
\mathfrak{D}_{2,2}=\beta_2(\alpha^2 p^2-\alpha\gamma p-\alpha p^2-2\alpha pq+\gamma q+pq+q^2),\\
\mathfrak{D}_{1,3}&=&-\beta_2 q(\gamma-2),\quad
\mathfrak{D}_{2,3}=\beta_2(2\alpha p+\delta q-\alpha\delta p-2q)
\end{eqnarray*}
is given by
\[
f(x)=c_1 H_c(a_1,a_2,a_3,a_4,a_5;x)+c_2 x^{1-\gamma}H_c(a_1,-a_2,a_3,a_4,a_5;x)+
\]
\[
e^{\widetilde{a}x}\left[c_3H_c(\widetilde{a}+a_1,a_2,a_3,a_4,a_5;x)+c_4 x^{1-\gamma}H_c(\widetilde{a}+a_1,-a_2,a_3,a_4,a_5;x)\right].
\]
It is interesting to observe that also in this case Maple $16$ cannot solve the fourth order ODE $Lf=0$ discussed above.
\item
The cases $\alpha=\gamma=\delta=2$ and $q=p=-2$, or $\gamma=\delta=2$, $p=0$, and $q=-2$, or $\delta=p=q=0$, and $\gamma=2$, or $\gamma=p=q=0$, and $\delta=2$, or $\gamma=p=q=\delta=0$ have been already discussed in Section~\ref{HeunN2}. 
\item
Case $\alpha=1$, $\gamma=2$, $q=p$, and $\delta=0$. The solutions to  $P_{CH}\psi=0$ and $Q\varphi=0$ have been already computed in Section~\ref{cHn1} and Section~\ref{HeunN2}, respectively. Finally, the solution to the equation $Lf=0$ with $L$ as in (\ref{LCHn2}) and non-vanishing coefficients $\Gamma=\beta_2 p+\beta_1$, $\mathfrak{A}_2=4\beta_2$, $\lambda=\beta_1 p+\beta_0$, $\mathfrak{B}_{1,1}=3\Gamma$, $\mathfrak{C}_{1,1}=2\Gamma$ and $\mathfrak{D}_{1,1}=\eta$
is
\[
f(x)=\frac{1}{x}\left(c_1+c_2 e^{-px}+c_3 e^{m_- x}+c_4 e^{m_+ x}\right)
\]
with $m_\pm$ defined in (\ref{mpmCH}).
\item
Case $\alpha=1$, $\gamma=q=0$, and $\delta=2$. The solutions to  $P_{CH}\psi=0$ and $Q\varphi=0$ have been already computed in Section~\ref{cHn1} and Section~\ref{HeunN2}, respectively. Moreover, the solution to the equation $Lf=0$ with $L$ as in (\ref{LCHn2}) and coefficients $\Gamma$, $\mathfrak{A}_2$ and $\lambda$ as in the case above and $\mathfrak{B}_{2,1}=3\Gamma$, $\mathfrak{C}_{2,1}=2\Gamma$ and $\mathfrak{D}_{2,1}=\eta$
is 
\[
f(x)=\frac{1}{x-1}\left(c_1+c_2 e^{-px}+c_3 e^{m_- x}+c_4 e^{m_+ x}\right).
\]
\item
Case $\alpha=\gamma=\delta=q=0$. The solutions to  $P_{CH}\psi=0$ and $Q\varphi=0$ have been already computed in Section~\ref{cHn1} and Section~\ref{HeunN2}, respectively. Moreover, the solution to the equation $Lf=0$ with $L$ as in (\ref{LCHn2}) and non-vanishing coefficients $\Gamma$, $\lambda$ and $\eta$ given as in the previous case is
\[
f(x)=c_1+c_2 e^{-px}+c_3 e^{m_- x}+c_4 e^{m_+ x}.
\] 
\end{enumerate}
In the case of the reduced confluent Heun differential operator $P_{RCH}$ we find that the most general second degree differential operator $Q$ semi-commuting with $P_{RCH}$ is given by 
\[
Q=\beta_2 d^2_x+\left[\beta_2\left(\frac{\gamma}{x}+\frac{\delta}{x-1}\right)+\beta_1\right]d_x+\frac{\beta_1\gamma-2\beta_2 q}{2x}+\frac{2\beta_2(k+q)}{2(x-1)}+\beta_0. 
\]
By means of (\ref{COR2}) we can verify that the operators $P_{RCH}$ and $Q$ will commute whenever $\gamma=q=\delta=2$ and $k=0$, or $\gamma=2$ and $k=q=\delta=0$, or $\gamma=k=q=0$ and $\delta=2$, or $\gamma=k=q=\delta=0$ but these cases reduce to one of the cases treated above.

\subsection{The biconfluent Heun operator}
\subsubsection{The case $n=1$}\label{BCHn1}
Using (\ref{COR1}) yields the following family of first degree operators   
\[
Q=\beta_1 d_x+\frac{\beta_1}{2}\left(\frac{\tau}{x}+\frac{\nu}{x^2}\right)+\beta_0
\]
semi-commuting with the operator $P_{BCH}$ associated to the biconfluent Heun equation. The commutativity condition (\ref{COR2}) will be satisfied for certain choices of the parameters that we list here below. Note that if $L:=QP_{BCH}=P_{BCH}Q$, the solutions to the equations $Q\varphi=0$ and $P_{BCH}\psi=0$ allow to construct the general solution to the third order ODE $Lf=0$ with
\begin{equation}\label{LBHn1}
L=\beta_1d^3_x+\left(K+\frac{A}{x}\right)d^2_x+\left(R+\frac{B}{x}\right)d_x+C+\frac{D}{x}.
\end{equation}
\begin{enumerate}
\item
Case $\nu=q=\tau=0$. The solution to $P_{BCH}\psi=0$ is $\psi(x)=c_1 e^{\alpha_+ x}+c_2 e^{\alpha_- x}$ with $\alpha_\pm=\left(1\pm\sqrt{1+4\alpha}\right)/2$. Moreover, the solution to $Q\varphi=0$ has been already obtained in Section~\ref{n1HH}. Finally, the general solution to the ODE $Lf=0$ with $L$ as in (\ref{LBHn1}) and non-vanishing coefficients $K=\beta_0-\beta_1$, $R=-(\beta_1\alpha+\beta_0)$ and $C=-\beta_0\alpha$ is
\[
f(x)=c_1 e^{\alpha_+ x}+c_2 e^{\alpha_- x}+c_3 e^{-\frac{\beta_0}{\beta_1}x}.
\]
\item
Case $\nu=0$, $q=1$, and $\tau=2$. The solution to $P_{BCH}\psi=0$ is $\psi(x)=\frac{1}{x}\left(c_1 e^{\alpha_+ x}+c_2 e^{\alpha_- x}\right)$ with $\alpha_\pm$ defined in the previous case. Furthermore, the solution to $Q\varphi=0$ has been obtained in Section~\ref{n1HH}. The general solution to the ODE $Lf=0$ with $L$ as in (\ref{LBHn1}), $K$ and $R$ as in the case above and $A=3\beta_1$, $B=2K$ and $D=R$ is
\[
f(x)=\frac{1}{x}\left(c_1 e^{\alpha_+ x}+c_2 e^{\alpha_- x}+c_3 e^{-\frac{\beta_0}{\beta_1}x}\right).
\]
\end{enumerate}

\subsubsection{The case $n=2$}\label{BCHn2}
By means of (\ref{COR3}) and (\ref{COR4}) we find the following family of second degree operators   
\[
Q=\beta_2 d^2_x+\left[\beta_2\left(\frac{\tau}{x}+\frac{\nu}{x^2}\right)+\beta_1\right]d_x+\frac{1}{2}\left[\frac{\tau(\beta_1+\beta_2)-2\beta_2 q}{x}+\frac{\nu(\beta_1+\beta_2)}{x^2}\right]+\beta_0 
\]
semi-commuting with the operator $P_{BCH}$ associated to the biconfluent Heun equation. The commutativity condition (\ref{COR5}) will be satisfied for certain choices of the parameters that we list here below. Notice that if $L:=QP_{BCH}=P_{BCH}Q$, the solutions to the equations $Q\varphi=0$ and $P_{BCH}\psi=0$ allow to construct the general solution to the fourth order ODE $Lf=0$ with
\begin{equation}\label{LBHn2}
L=d^4_x+\sum_{n=0}^2\frac{A_n}{x^n} d^3_x+\sum_{n=0}^4\frac{B_n}{x^n}d^2_x+\sum_{n=0}^5\frac{C_n}{x^n}d_x+\sum_{n=0}^4\frac{D_n}{x^n}.
\end{equation}
\begin{enumerate}
\item
Case $\beta_1=-\beta_2$. The solution to the equation $P_{BCH}\psi=0$ is
\[
\psi(x)=x^{\frac{1}{2}(1-\tau)}\left[c_1 e^{\rho_{-}(x)} H_D(b_1,b_2,b_3,b_4;\omega(x))+c_2 e^{\rho_{+}(x)} H_D(-b_1,b_2,b_3,b_4;\omega(x))\right],
\]
where $H_D$ denotes the double confluent Heun function \cite{Ronveaux},
\[
\rho_{-}(x)=\frac{1}{2}[1-\epsilon(\nu)]\sqrt{4\alpha+1}x,\quad 
\rho_{+}(x)=\frac{2\nu+[1+\epsilon(\nu)\sqrt{4\alpha+1}]x^2}{2x},\quad \omega(x)=\frac{ix\sqrt[4]{(4\alpha+1)\nu^2}+\nu}{ix\sqrt[4]{(4\alpha+1)\nu^2}-\nu},
\]
and
\begin{eqnarray*}
b_1&=&4i\sqrt[4]{(4\alpha+1)\nu^2},\\
b_2&=&\frac{2i\epsilon(\nu)(2q-\tau)\sqrt[4]{(4\alpha+1)^3\nu^2}-2(4\alpha+1)\left[i(\tau-2)\sqrt[4]{(4\alpha+1)\nu^2}+\frac{\tau^2}{2}-\nu-\tau+|\nu|\sqrt{4\alpha+1}+\frac{1}{2}\right]}{4\alpha+1},\\
b_3&=&\frac{4i\sqrt[4]{(4\alpha+1)\nu^2}\left[\epsilon(\nu)(2q-\tau)\sqrt{4\alpha+1}+(\tau-2)(4\alpha+1)\right]}{4\alpha+1},\\
b_4&=&\frac{2i\epsilon(\nu)(2q-\tau)\sqrt[4]{(4\alpha+1)^3\nu^2}-2(4\alpha+1)\left[i(\tau-2)\sqrt[4]{(4\alpha+1)\nu^2}-\frac{\tau^2}{2}+\nu+\tau-|\nu|\sqrt{4\alpha+1}-\frac{1}{2}\right]}{4\alpha+1}.
\end{eqnarray*}
Here, $\epsilon(\cdot)$ denotes the sign function. If we let $\mu=\beta_0/\beta_2$, the solution to the equation $Q\varphi=0$ is given by
\[
\varphi(x)=x^{\frac{1}{2}(\tau-1)}\left[c_3 e^{\sigma_{-}(x)} H_D(\ell_1,\ell_2,\ell_3,\ell_4;r(x))+c_2 e^{\sigma_{+}(x)} H_D(-\ell_1,\ell_2,\ell_3,\ell_4,\ell_5;r(x))\right],
\]
with $\sigma_{-}$, $\sigma_+$ and $r$ formally given by $\rho_{-}$, $\rho_{+}$ and $\omega$ with $\alpha$ replaced by $-\mu$ and
\begin{eqnarray*}
\ell_1&=&4i\sqrt[4]{(1-4\mu)\nu^2},\\
\ell_2&=&\frac{2i\epsilon(\nu)(\tau-2q)\sqrt[4]{(1-4\mu)^3\nu^2}+2(1-4\mu)\left[i(\tau-2)\sqrt[4]{(1-4\mu)\nu^2}+\frac{\tau^2}{2}-\nu-\tau+|\nu|\sqrt{1-4\mu}+\frac{1}{2}\right]}{4\mu-1},\\
\ell_3&=&\frac{4i\sqrt[4]{(1-4\mu)\nu^2}\left[\epsilon(\nu)(2q-\tau)\sqrt{1-4\mu}+(\tau-2)(1-4\mu)\right]}{1-4\mu},\\
\ell_4&=&\frac{2i\epsilon(\nu)(\tau-2q)\sqrt[4]{(1-4\mu)^3\nu^2}+2(1-4\mu)\left[i(\tau-2)\sqrt[4]{(1-4\mu)\nu^2}-\frac{\tau^2}{2}+\nu+\tau-|\nu|\sqrt{1-4\mu}-\frac{1}{2}\right]}{4\mu-1}.
\end{eqnarray*} 
Then, the general solution to the equation $Lf=0$ with $L$ as in (\ref{LBHn2}) and non-vanishing coefficients
\begin{eqnarray*}
A_0&=&-2,\quad A_1=2\tau,\quad A_2=2\nu, B_0=1+\mu-\alpha,\quad B_1=-2(\tau+q),\quad B_2=\tau(\tau-2)-2\nu,\\ B_3&=&2\nu(\tau-2),\quad B_4=\nu^2,\quad
C_0=\alpha-\mu,\quad C_1=(\mu-\alpha)\tau+2q,\quad C_2=(\mu-\alpha)\nu+(1-2q)\tau+2q,\\ C_3&=&-[\tau(\tau-2)+2\nu(q-1)],\quad C_4=-3\nu(\tau-2),\quad C_5=-2\nu^2,\\
D_0&=&-\mu\alpha,\quad D_1=q(\alpha-\mu),\quad D_2=q(q-1),\quad D_3=q(\tau-2),\quad D_4=q\nu\\
\end{eqnarray*}
is
\[
f(x)=x^{\frac{1}{2}(1-\tau)}\left[c_1 e^{\rho_{-}(x)} H_D(b_1,b_2,b_3,b_4;\omega(x))+c_2 e^{\rho_{+}(x)} H_D(-b_1,b_2,b_3,b_4;\omega(x))\right]+
\]
\[
x^{\frac{1}{2}(\tau-1)}\left[c_3 e^{\sigma_{-}(x)} H_D(\ell_1,\ell_2,\ell_3,\ell_4;r(x))+c_2 e^{\sigma_{+}(x)} H_D(-\ell_1,\ell_2,\ell_3,\ell_4,\ell_5;r(x))\right].
\]
Also in this case Maple $16$ is not able to solve the above fourth order ODE.
\item
Case $\nu=q=\tau=0$. The solution to $P_{BCH}\psi=0$ is given in case $1.$ ($n=1$) of the biconfluent Heun operator. Moreover, the solution to the equation $Q\varphi=0$ has been already studied in Section`\ref{n1HH}. Finally, the solution of to the ODE $Lf=0$ with with $L$ as in (\ref{LBHn1}) and non-vanishing coefficients $A_0=\beta_1-\beta_2$, $B_0=\beta_0-\beta_1-\beta_2\alpha$, $C_0=-\beta_0-\beta_1\alpha$ and $D_0=-\beta_0\alpha$ is
\[
f(x)=c_1 e^{\alpha_+ x}+c_2 e^{\alpha_- x}+c_3 e^{m_- x}+c_4 e^{m_- x},
\]
where $m_\pm$ and $\alpha_\pm$ have been defined in (\ref{mpmCH}) and in case 1 Section~\ref{BCHn1}, respectively.
\item
Case $\nu=0$, $q=1$, and $\tau=2$. The solution to $P_{BCH}\psi=0$ is given as in case $2.$ ($n=1$) of the biconfluent Heun operator. Moreover, the solution to $Q\varphi=0$ has been studied in Section~\ref{HeunN2}. Finally, the general solution to the ODE $Lf=0$ with $L$ as in (\ref{LBHn1}) and non-vanishing coefficients $A_0$, $B_0$, $C_0$, $D_0$ given in the case above and $A_1=4\beta_2$, $B_1=3A_0$, $C_1=-2B_0$ and $D_1=C_0$ is
\[
f(x)=\frac{1}{x}\left(c_1 e^{\alpha_+ x}+c_2 e^{\alpha_- x}+c_3 e^{m_- x}+c_4 e^{m_- x}\right),
\]
where $m_\pm$ and $\alpha_\pm$ have been defined in (\ref{mpmCH}) and in case 1 Section~\ref{BCHn1}, respectively.
\end{enumerate}

\subsection{The double confluent Heun operator}
\subsubsection{The case $n=1$}\label{DCHHn1}
Using (\ref{COR1}) yields the same family of first degree operators $Q$ obtained for the biconfluent Heun operator in the case $n=1$. The commutativity condition (\ref{COR2}) will be satisfied for $\alpha=\tau/2$, $\nu=0$, and $q=(\tau/2)-(\tau^2/4)$. For this choice of the parameters we have
\[
P_{DCH}=d^2_x+\left(\frac{\tau}{x}-1\right)d_x+\frac{\tau(\tau-2-2x)}{4x^2},\quad
Q=\beta_1 d_x+\frac{\beta_1\tau}{2x}d_x+\beta_0.
\]
Then, the operator $L:=QP_{DCH}=P_{DCH}Q$ is of the form
\[
L=\beta_1 d^3_x+\sum_{n=0}^1\frac{a_n}{x^n}d^2_x+\sum_{n=0}^2\frac{b_n}{x^n}d_x+\sum_{n=0}^2\frac{c_n}{x^n}
\]
with $a_0=\beta_0-\beta_1$, $a_1=3\beta_1\tau/2$, $b_0=-\beta_0$, $b_1=\tau a_0$, $b_2=3\beta_1\tau(\tau-2)$, $c_1=b_0\tau$ and $c_2=a_0\tau(\tau-2)$ and the solution to the ODE $Lf=0$ reads
\[
f(x)=x^{-\frac{\tau}{2}}\left[c_1+c_2 e^x+c_3 e^{-\frac{\beta_0}{\beta_1}x}\right].
\]
 
\subsubsection{The case $n=2$}
By means of (\ref{COR3}) and (\ref{COR4}) we find the following  family of second degree operators  
\[
Q=\beta_2 d^2_x+\left[\beta_2\left(\frac{\tau}{x}+\frac{\nu}{x^2}\right)+\beta_1\right]d_x+\frac{\beta_1\tau+\beta_2(\tau-2\alpha)}{2x}+\frac{\beta_1\nu+\beta_2(\nu-2q)}{2x^2}+\beta_0
\]
semi-commuting with $P_{DCH}$. The commutativity condition (\ref{COR5}) will be satisfied for certain choices of the parameters that we list here below. Notice that if $L:=QP_{DCH}=P_{DCH}Q$, the solutions to the equations $Q\varphi=0$ and $P_{DCH}\psi=0$ allow to construct the general solution to the fourth order ODE $Lf=0$ with
\begin{equation}\label{LDCHn2}
L=d^4_x+\sum_{n=0}^2\frac{A_n}{x^n} d^3_x+\sum_{n=0}^4\frac{B_n}{x^n}d^2_x+\sum_{n=0}^5\frac{C_n}{x^n}d_x+\sum_{n=0}^5\frac{D_n}{x^n}.
\end{equation}
\begin{enumerate}
\item
Case $\beta_1=-\beta_2$. The solution to the equation $P_{DCH}\psi=0$ is
\[
\psi(x)=x^{\frac{1}{2}(1-\tau)}\left[c_1 e^{x} H_D\left(m_1,m_2,m_3,m_4;\frac{x-\sqrt{\nu}}{x+\sqrt{\nu}}\right)+c_2 e^{\frac{x}{\nu}} H_D\left(-m_1,m_2,m_3,m_4;\frac{x-\sqrt{\nu}}{x+\sqrt{\nu}}\right)\right],
\]
where $H_D$ denotes the double confluent Heun function \cite{Ronveaux,Slavyanov} and $m_1=-4\sqrt{\nu}$, $m_2=4(\alpha-1)\sqrt{\nu}-\tau^2-4q+4\nu+2\tau-1$, $m_3=8\sqrt{\nu}(\alpha-\tau+1)$, $m_4=8(\alpha-1)\sqrt{\nu}-m_2$. Furthermore, the solution to the equation $Q\varphi=0$ is given by
\[
\varphi(x)=x^{\frac{1}{2}(1-\tau)}\left[c_3 e^{\sigma_{-}(x)} H_D(t_1,t_2,t_3,t_4;r(x))+c_2 e^{\sigma_{+}(x)} H_D(-t_1,t_2,t_3,t_4,t_5;r(x))\right],
\]
where the functions $\sigma_\pm(\cdot)$ and $r(\cdot)$ have been defined in Section~\ref{BCHn2} and
\begin{eqnarray*}
t_1&=&4i\sqrt[4]{(1-4\mu)\nu^2},\\
t_2&=&\frac{2i\epsilon(\nu)(\tau-2\alpha)\sqrt[4]{(1-4\mu)^3\nu^2}+2(1-4\mu)\left[i(\tau-2)\sqrt[4]{(1-4\mu)\nu^2}+\frac{\tau^2}{2}-\nu-\tau+2q+|\nu|\sqrt{1-4\mu}+\frac{1}{2}\right]}{4\mu-1},\\
t_3&=&\frac{4i\sqrt[4]{(1-4\mu)\nu^2}\left[\epsilon(\nu)(\tau-2\alpha)\sqrt{1-4\mu}+(\tau-2)(4\mu-1)\right]}{4\mu-1},\\
t_4&=&\frac{2i\epsilon(\nu)(\tau-2\alpha)\sqrt[4]{(1-4\mu)^3\nu^2}+2(1-4\mu)\left[i(\tau-2)\sqrt[4]{(1-4\mu)\nu^2}-\frac{\tau^2}{2}+\nu+\tau-2q-|\nu|\sqrt{1-4\mu}-\frac{1}{2}\right]}{4\mu-1}.
\end{eqnarray*} 
Finally, the general solution to the ODE $Lf=0$ with $L$ as in (\ref{LDCHn2}) and non-vanishing coefficients
\begin{eqnarray*}
A_0&=&-2,\quad A_1=2\tau,\quad A_2=2\nu,\quad
B_0=1+\mu,\quad B_1=-2(\tau+\alpha),\quad B_2=\tau(\tau-2)-2\nu-2q,\\
B_3&=&2\nu(\tau-2),\quad B_4=\nu^2,\quad
C_0=-\mu,\quad C_1=2\alpha+\mu\tau,\quad C_2=\tau(1-\alpha)+2\alpha+2q+\mu\nu,\\ C_3&=&-[2(\nu-\alpha)+(2-\tau)(2q+\tau)],\quad
C_4=\nu(6-3\tau-2q),\quad C_5=-2\nu^2,\quad
D_1=-\mu\alpha,\\
D_2&=&\alpha(\alpha-1)-\mu q,\quad D_3=2\alpha q+\alpha\tau-2\alpha-2q,\quad D_4=\alpha\nu+q^2+2q\tau-6q,\quad D_5=2q\nu
\end{eqnarray*}
is
\[
f(x)=x^{\frac{1}{2}(1-\tau)}\left\{c_1 e^{x} H_D\left(m_1,m_2,m_3,m_4;\frac{x-\sqrt{\nu}}{x+\sqrt{\nu}}\right)+c_2 e^{\frac{x}{\nu}} H_D\left(-m_1,m_2,m_3,m_4;\frac{x-\sqrt{\nu}}{x+\sqrt{\nu}}\right)\right.+
\]
\[
\left.c_3 e^{\sigma_{-}(x)} H_D(t_1,t_2,t_3,t_4;r(x))+c_2 e^{\sigma_{+}(x)} H_D(-t_1,t_2,t_3,t_4,t_5;r(x))\right\}.
\]
Also in this case Maple $16$ fails to find the general solution to  the ODE $Lf=0$.
\item
Case $\alpha=\tau/2$, $\nu=0$, $q=(\tau/2)-(\tau/2)^2$. By means of the results obtained in Section~\ref{DCHHn1} the general solution to the ODE $Lf=0$ with $L$ as in (\ref{LDCHn2}) and non-vanishing coefficients 
\begin{eqnarray*}
A_0&=&\beta_1-\beta_2,\quad A_1=2\beta_2\tau,\quad B_0=\beta_0-\beta_1,\quad B_1=\frac{3}{2}A_0\tau,\quad B_2=\frac{3}{2}\beta_2\tau(\tau-2),\quad C_0=-\beta_0,\quad C_1=B_0 \tau,\\
C_2&=&\frac{3}{4}A_0\tau(\tau-2),\quad C_3=\beta_2\tau(\tau-2)(\tau-4),\quad D_1=-\frac{\beta_0\tau}{2},\quad D_2=\frac{B_0}{4}\tau(\tau-2),\quad D_3=\frac{A_0}{8}\tau(\tau-2)(\tau-4),\\
D_4&=&\frac{\beta_2}{16}\tau(\tau-2)(\tau-4)(\tau-16)
\end{eqnarray*}
is
\[
f(x)=x^{-\frac{\tau}{2}}\left[c_1+c_2 e^x+c_3 e^{m_- x}+c_4 e^{m_+ x}\right]
\]
where $m_\pm$ has been defined in (\ref{mpmCH}).
\end{enumerate}

\subsection{The triconfluent Heun operator}

\subsubsection{The case $n=1$}
Let $P_{TCH}$ and $P_{RTCH}$ be the operators associated to the triconfluent Heun equation and  the reduced triconfluent Heun equation, respectively. Using (\ref{COR1}) yields the following family of first degree operators   
\[
Q_1=\beta_1 d_x-\frac{\beta_1}{2}x^2+\beta_0,\quad
Q_2=\beta_1 d_x+\beta_0
\]
semi-commuting with the operators $P_{TCH}$ and $P_{RTCH}$, respectively. In both cases, the commutativity condition (\ref{COR2}) will be satisfied whenever $\beta_1=0$ but then, $Q_1$ and $Q_2$ become trivial and therefore, there are no nontrivial first degree operators commuting with $P_{TCH}$ or $P_{RTCH}$.

\subsubsection{The case $n=2$}
By means of (\ref{COR3}) and (\ref{COR4}) we find the following family of second degree operators   
\[
Q=\beta_2 d^2_x+(\beta_1-\beta_2 x^2) d_x+\frac{1}{2}(\sigma\beta_2-\beta_1)x^2+\beta_2\alpha x+\beta_0
\]
semi-commuting with the operator $P_{TCH}$. The commutativity condition (\ref{COR5}) will be satisfied whenever $\beta_1=\sigma\beta_2$. Furthermore, the solution to the equation $P_{TCH}\psi=0$ is
\[
\psi(x)=c_1 H_T(-\sqrt[3]{9}q,3\alpha+3,-\sqrt[3]{3}\sigma;x/\sqrt[3]{3})+c_2 e^{\frac{x^3}{3}-\sigma x}H_T(-\sqrt[3]{9}q,-3\alpha-3,-\sqrt[3]{3}\sigma;-x/\sqrt[3]{3}),
\]
where $H_T$ denotes the triconfluent Heun function \cite{Ronveaux} and the solution to $Q\varphi=0$ is given by
\[
\varphi(x)=c_3 H_T(-\sqrt[3]{9}\mu,3\alpha+3,-\sqrt[3]{3}\sigma;x/\sqrt[3]{3})+c_4 e^{\frac{x^3}{3}-\sigma x}H_T(\sqrt[3]{9}\mu,-3\alpha-3,-\sqrt[3]{3}\sigma;-x/\sqrt[3]{3}),\quad\mu=\frac{\beta_0}{\beta_2}.
\]
Then, the general solution to the fourth order equation $Lf=0$ with
\[
L=d^4_x+A_3(x)d_x^3+A_2(x)d^2_x+A_1(x)d_x+A_0(x),
\]
where
\begin{eqnarray*}
A_3(x)&=&2(\sigma-x^2),\quad
A_2(x)=x^4-2\sigma x^2+(2\alpha-4)x+\sigma^2-q+\mu,\\
A_1(x)&=&2(1-\alpha)x^3+(q-\mu)x^2+2\sigma(\alpha-1)x+\sigma(q-\mu)+2(\alpha-1),\quad
A_0(x)=\alpha(1-\alpha)x^2+\alpha(q-\mu)x+\mu q-\alpha\sigma
\end{eqnarray*}
is $f(x)=\psi(x)+\varphi(x)$. Also in this case Maple $16$ is not able to solve the ODE $Lf=0$. We conclude by observing that the following family of second degree operators   
\[
Q=\beta_2 d^2_x+\beta_1 d_x+\beta_2\left(A_1 x+A_2 x^2-\frac{9}{4}x^4\right)+\beta_0
\]
semi-commutes with the operator $P_{RTCH}$. The commutativity condition (\ref{COR5}) will be satisfied for $\beta_1=0$. Then, the general solution to the equation $P_{RTCH}\psi=0$ is
\begin{equation}\label{psiRTC}
\psi(x)=c_1 e^{-\frac{x^3}{2}+\frac{A_2}{3}x} H_T\left(A_0+\frac{A_2^2}{9},A_1,-\frac{2}{3}A_2;x\right)+c_2 e^{\frac{x^3}{2}-\frac{A_2}{3}x}H_T\left(A_0+\frac{A_2^2}{9},-A_1,-\frac{2}{3}A_2;-x\right),
\end{equation}
and the general solution to $Q\varphi=0$ is given by
\begin{equation}\label{phiRTC}
\varphi(x)=c_3 e^{-\frac{x^3}{2}+\frac{A_2}{3}x} H_T\left(\mu+\frac{A_2^2}{9},A_1,-\frac{2}{3}A_2;x\right)+c_4 e^{\frac{x^3}{2}-\frac{A_2}{3}x}H_T\left(\mu+\frac{A_2^2}{9},-A_1,-\frac{2}{3}A_2;-x\right)
\end{equation}
with $\mu$ defined as above. Hence, the general solution to the fourth order equation $Lf=0$ with
\[
L=d^4_x+B_2(x)d^2_x+B_1(x)d_x+B_0(x),
\]
where
\begin{eqnarray*}
B_2(x)&=&-\frac{9}{2}x^4+2A_2 x^2+2A_1 x+\mu+A_0,\quad
B_1(x)=2(-9x^3+2A_2 x+A_1),\\
B_0(x)&=&\frac{81}{16}x^8-\frac{9}{2}A_2 x^6-\frac{9}{2}A_1 x^5+\left[A_2^2-\frac{9}{4}(\mu+A_0)\right]x^4+2A_1 A_2 x^3+\left[A_1^2+A_2(A_0+\mu)-27\right]x^2+\\
&&A_1(A_0+\mu)x+A_0\mu+2A_2.
\end{eqnarray*}
is simply $f(x)=\psi(x)+\varphi(x)$. Also in this case Maple $16$ fails to find the general solution to the ODE $Lf=0$.

\section{Conclusions}
The study of the Heun equation, its generalizations and solutions is one of the next challenges in the area differential equations connected to mathematical physics. Being a generalization of the hypergeometric equation it is not surprising that its appearance in (mathematical) physics is manifold. A quite exhaustive review of its applications in physics and quantum chemistry is offered by  \cite{Ronveaux}. More recent and unexpected applications of the Heun family of equations in physics can be found in \cite{Slavyanov, Hortascu, Fiziev}. In general relativity the Heun equation has a well defined connection to the black hole physics \cite{BSW,CKTSK,BS}
and is used in quantum mechanics as well \cite{Bay, Tolstikhan, Hall}. In particular, one can connect it in an elegant way
to the Schr\"odinger equation \cite{U1, U2, U3}, the Stark effect \cite{Epstein} and other specific quantum mechanical problems \cite{BIWTRL}. 

From a mathematical point of view the study of the Heun equation and its confluent forms is far from being complete. Even though we can easily construct pairs of Frobenius series solutions around each  regular singular point and derive the asymptotics of the solutions for the point at infinity, the so-called connection problem for these local solutions is still open mainly due to the fact that the construction of integral representations for the solutions results to be a very difficult task. Some steps towards the resolution of this problem have being taken in \cite{RD} where a modification of the methods used in \cite{SS} allowed to solve the two-point connection problem for a subclass of the Heun equation. Other attempts can be found in \cite{Ronveaux,Slavyanov,PF}. 

Here, we have taken a slightly different point of view. We considered the Heun family of differential operators given in Table~\ref{Heun} and for each member of this family we constructed the corresponding most general class of commuting differential operators of degree one and degree two. This in turn allowed to show that using this method solutions of complicated higher order linear homogeneous differential equations with variable coefficients can be found analytically even though the software package Maple $16$ was not able to compute them.


\begin{thebibliography}{999}  
\bibitem{FWS}
G. Floquet, {\it{Sur la th$\acute{\mbox{e}}$orie des $\acute{\mbox{e}}$quations diff$\acute{\mbox{e}}$rentielles lin$\acute{\mbox{e}}$aires}}, Ann. Sci. $\acute{\mbox{E}}$cole Norm. Sup. {\bf{8}} (1879), suppl., 1; G. Wallenberg, {\it{\"{U}ber die Vertauschbarkeit homogener linearer Differentialausdr\"{u}cke}}, Arc. Math. Phys. {\bf{4}} (1903), 252; J. Schur, {\it{\"{U}ber vertauschbare lineare Differentialausdr\"{u}cke}}, Sitzungsber. der Berliner Math. Gesell. {\bf{4}} (1905), 2
\bibitem{BC123}
J. L. Burchnall and T. W. Chaundy, {\it{Commutative Ordinary Differential Operators}}, Proc. Royal Soc. London {\bf{21}} (1922), 420; ibidem {\bf{118}} (1928), 557; ibidem {\bf{134}} (1931), 471
\bibitem{Gorder}
R. A. Van Gorder, {\it{Computing semi-commuting differential operators in one and multiple variables}}, Math. Comm. {\bf{19}} (2014), 201
\bibitem{Heun}
K. Heun, {\it{Zur Theorie der Riemann'schen Functionen zweiter Ordnung mit vier Verzweigungspunkten}}, Math. Annalen {\bf{33}} (1889), 161
\bibitem{Ronveaux}
A. Ronveaux, {\it {Heun's Differential Equations}}, Oxford University Press, 1995
\bibitem{Slavyanov}
S. Ya. Slavyanov and W. Lay, {\it {Special Functions: A Unified Approach Based on Singularities}}, Oxford University Press, 2001
\bibitem{SS}
R. Sch\"{a}fke and D. Schmidt, {\it{The Connection Problem for General Linear Ordinary Differential Equations at Two Regular Singular Points with Applications in the Theory of Special Functions}}, SIAM J. Math. Anal. {\bf{11}} (1980), 848
\bibitem{BSW}
D. Batic, H. Schmid and M. Winklmeier, {\it{The generalized Heun equation in QFT in curved space-times}}, J. Phys. A{\bf{39}} (2006), 12559
\bibitem{Hortascu}
M. Hortascu, {\it{Heun Functions and their uses in Physics}}, Proceedings of the $13^{th}$ Regional Conference on Mathematical Physics, Antalya, Turkey, 27-31 October 2010, ed. by U. Camci and I. Semiz, pp. 23, World Scientific (2013)
\bibitem{Fiziev}
P. P. Fiziev, {\it{Heun as modern powerful tool for research in different scientific domains}}, arxiv: 1512.04015 [math-ph]
\bibitem{CKTSK}
S. Chandrasekhar, {\it The Mathematical Theory of Black Holes}, Oxford University Press, 1983; R. P. Kerr, {\it{Gravitational Field of a Spinning Mass as an Example of Algebraically Special Metrics}}, Phys. Rev. Lett. {\bf 11} (1963), 237; S. A. Teukolsky, {\it{Rotating black holes - separable wave equations for gravitational and electromagnetic perturbations}}, Phys. Rev. Lett. {\bf 29} (1972), 1114; H. Suzuki, E. Tagasugi and H. Umetsu, {\it{Perturbations of Kerr-de Sitter Black Holes and Heun's Equations}}, Prog. Theor. Phys. {\bf 100} (1998), 49; E. G. Kalnins et al, Proceedings of the international workshop  special functions, Hong Kong, 21-25 June 1999, ed. by C. Dunkl, M. Ismail and R. Wong, pp. 438, World Scientific (2000) 
\bibitem{BS}
D. Batic and H. Schmid, {\it{Heun equation, Teukolsky equation, and type-D metrics}}, J. Math. Phys. {\bf{48}} (2007), 042502-29
\bibitem{Bay}
K. Bay, W. Lay and A. Akopyan, {\it{Avoided crossings of the quartic oscillator}}, J. Phys.{\bf A30} (1997), 3057
\bibitem{Tolstikhan}
U. I. Tolstikhin and M. Matsuzawa, {\it{Hyperspherical elliptic harmonics and their relation to the Heun equation}}, Phys. Rev. {\bf A63} (2001),  032510
\bibitem{Hall}
R. L. Hall, N. Saad and K. D. Sen, {\it{Soft-core Coulomb potential and Heun's differential equation}}, J. Math. Phys. {\bf 51} (2010),  022107
\bibitem{U1}
D. Batic, R. Williams and M. Nowakowski, {\it{Potentials of the Heun class}}, J. Phys. {\bf A46} (2013), 245204  
\bibitem{U2}
D. Batic, D. Mills-Howell and M. Nowakowski, {\it{Potentials of the Heun class: The triconfluent case}} J. Math Phys. {\bf 56} (2015), 052106
\bibitem{U3}
G. Natanson, {\it Heun-Polynomial Representation of Regular-at-Infinity Solutions for the Basic SUSY Ladder of Hyperbolic P\"oschl-Teller 
Potentials Starting from the Reflectionless Symmetric Potential Well}, arxiv:1410.1515 [math-ph]
\bibitem{Epstein}
P. S. Epstein, {\it{The Stark Effect from the Point of View of Schroedinger's Quantum Theory}}, Phys. Rev. {\bf 2} (1926), 695
\bibitem{BIWTRL}
V. Balan et al., {\it     	
Confluent Heun functions and the Coulomb problem for spin 1/2 particle in Minkowski space}, arxiv:1410.8300 [math-ph]; T. A. Ishkhanyan et al., {\it{Expansions of the solutions of the biconfluent Heun equation in terms of incomplete Beta and Gamma functions}},  J.Contemp.Phys. {\bf 51} (2016), 229; A. H. Wilson, {\it{A Generalised Spheroidal Wave Equation}}, Proc. Roy. Soc. London {\bf A118} (1928), 617; T. T. Truong and D. Bazzali, {\it{Exact low-lying states of two interacting equally charged particles in a magnetic field}}, Phys. Lett. {\bf A269} (2000), 186; A. Ralko and T. T. Truong, {\it{Heun functions and the energy spectrum of a charged particle on a sphere under a magnetic field and Coulomb force}}, J. Phys. {\bf A35} (2002), 9671; E. W. Leaver, {\it{Solutions to a generalized spheroidal wave equation: Teukolsky's equations in general relativity, and the two?center problem in molecular quantum mechanics}}, J. Math Phys. {\bf 27} (1986), 1238
\bibitem{RD}
R. Williams and D. Batic, {\it{The Two-Point Connection Problem for a Sub-Class of the Heun Equation}}, Journal of Inequalities and Special Functions {\bf{6}} (2015), 1
\bibitem{PF}
P. P. Fiziev, {\it{A new approach to the connection problem for local solutions to the general Heun equation}}, arXiv:1606.08539 [math-ph]
\end{thebibliography}
\end{document}